\documentclass[a4paper,12pt]{article}
\usepackage{amsfonts}

\usepackage{amsmath}
\usepackage[utf8]{inputenc}
\usepackage{fullpage}
\usepackage{boxedminipage}
\usepackage{listings}
\usepackage{minitoc}
\usepackage[pdftex]{graphicx}
\usepackage{graphicx}

\makeatletter \@addtoreset{equation}{section} \makeatother

\newif\ifpdf \ifx\pdfoutput\undefined \pdffalse
\else \pdfoutput=1 \pdftrue \fi \ifpdf \else \fi

\begin{document}

\ifpdf \DeclareGraphicsExtensions{.pdf, .jpg, .tif} \else %
\DeclareGraphicsExtensions{.eps, .jpg} \fi
\begin{titlepage}

    \thispagestyle{empty}
    \begin{flushright}
        \hfill{CERN-PH-TH/2006-027} \\
        \hfill{\texttt{hep-th/0602161}}
    \end{flushright}

    \vspace{35pt}
    \begin{center}
        { \LARGE{\bf On some properties of\\ \vspace{10pt} the Attractor Equations }}

        \vspace{60pt}

        {\bf Stefano Bellucci$^\clubsuit$, Sergio Ferrara$^{\diamondsuit\clubsuit}$ and \ Alessio Marrani$^{\spadesuit\clubsuit}$}

        \vspace{30pt}

        {$\clubsuit$ \it INFN,\\
        Laboratori Nazionali di Frascati, \\
        Via Enrico Fermi 40,00044 Frascati, Italy\\
        \texttt{bellucci,marrani@lnf.infn.it}}

        \vspace{15pt}

        {$\diamondsuit$ \it Physics Department,\\
        Theory Unit, CERN, \\
        CH 1211, Geneva 23, Switzerland\\
        \texttt{sergio.ferrara@cern.ch}}

        \vspace{15pt}

        {$\spadesuit$ \it Museo Storico della Fisica e\\
        Centro Studi e Ricerche ``Enrico Fermi"\\
        Via Panisperna 89A, Compendio Viminale,\\00184 Roma, Italy}

        \vspace{15pt}

        \vspace{20pt}

        {ABSTRACT}
    \end{center}

    \vspace{10pt}

    We discuss the Attractor Equations of $N=2$, ${d=4}$
    supergravity in an extremal black hole background with arbitrary
    electric and magnetic fluxes (charges) for field-strength
    two-forms.

    The effective one-dimensional Lagrangian in the radial
    (evolution) variable exhibits features of a spontaneously broken
    supergravity theory. Indeed, non-BPS Attractor solutions correspond
    to the vanishing determinant of a (fermionic) gaugino mass
    matrix. The stability of these solutions is controlled by the
    data of the underlying Special K\"{a}hler Geometry of the vector
    multiplets' moduli space.

    Finally, after analyzing the 1-modulus case more in detail, we
    briefly comment on the choice of the K\"{a}hler gauge and its
    relevance for the recently discussed entropic functional.
    \vspace{150pt}

\end{titlepage}
\newpage \baselineskip 6 mm

\section{Introduction}

It is known that the Attractor Equations for stationary, spherically
symmetric, asymptotically flat black holes are described by an effective
one-dimensional Lagrangian (\cite{1}, \cite{2}), with an effective scalar
potential and effective fermionic ``mass terms'' terms controlled by the
field-strength fluxes, i.e. electric and magnetic charges
\begin{equation}
e_{\Lambda }\equiv \int_{S_{\infty }^{2}}\mathcal{G}_{\Lambda
},~~~~~m^{\Lambda }\equiv \int_{S_{\infty }^{2}}\mathcal{F}^{\Lambda
},~~~~~~\Lambda =0,1,...,n_{V},
\end{equation}
where, in the case of $N=2$, $d=4$ supergravity, $n_{V}$ denotes the number
of Abelian vector supermultiplets. Here $\mathcal{F}^{\Lambda }=dA^{\Lambda
} $ and $\mathcal{G}_{\Lambda }$ is the ``dual'' field-strength two-form (
\cite{3}, \cite{4}).

The ``apparent'' gravitino mass is given by the central charge function $%
Z\left( z,\overline{z};e,m\right) $, where $z^{i}$ ($i=1,...,n_{V}$) are the
complex coordinates of the Special K\"{a}hler-Hodge manifold $\mathcal{V}$ ($%
dim_{\mathbb{C}}\mathcal{V}=n_{V}$) of vector supermultiplets. The gaugino
mass matrix $\Lambda _{ij}$ is given by \cite{5}
\begin{equation}
\Lambda _{ij}=C_{ijk}g^{k\overline{k}}\overline{D}_{\overline{k}}\overline{Z}%
,
\end{equation}
where $C_{ijk}$ is the (K\"{a}hler-covariantly holomorphic) completely
symmetric, rank-3 tensor of Special K\"{a}hler Geometry (SKG), and it should
also be recalled that $\overline{D}_{\overline{i}}Z=0$. Finally, the
supersymmetry order parameter, related to the mixed gravitino-gaugino
coupling, is given by $D_{i}Z=\left( \partial _{i}+\frac{1}{2}\partial
_{i}K\right) Z$, with $K$ being the K\"{a}hler potential in $\mathcal{V}$.

In this paper we intend to discuss some general features of the Attractor
Equations \cite{1}, relating extremal non-BPS black holes exhibiting an
Attractor behavior to a degenerate matrix. Such a matrix must have vanishing
determinant in order for a non-BPS, extremal black hole with Attractor
behavior to exist.

The plan of the paper is as follows. In Sect. 2 we recall the Attractor
potential for a generic $N=2$, $d=4$ supergravity theory and its BPS
Attractor solutions (\cite{6}, \cite{7}, \cite{8}). Then, in Sect. 3 we
discuss the non-BPS case and the related eigenvalue problem, with an
explicit application to the one-modulus case. In Sect. 4 we give the Hessian
matrix which controls the stability of the non-BPS extrema and discuss some
properties, especially on the one-modulus case. Finally, in Sect. 5 we try
to connect our results to the entropic functional recently introduced in
\cite{GSV} in connection with the conjectured relation between the black
hole entropy and the topological string partition function \cite{OSV1}.

\section{Black Hole Potential and Supersymmetric Attractors}

In $N=2$, $d=4$ supergravity the general form of the ``effective black hole
potential'' function reads (\cite{1}, \cite{4}, \cite{8})
\begin{equation}
V_{BH}\left( z,\overline{z};e,m\right) =\left| Z\right| ^{2}+g^{i\overline{i}%
}\left( D_{i}Z\right) \left( \overline{D}_{\overline{i}}\overline{Z}\right) .
\label{VBH}
\end{equation}
The aim of the present work is to discuss the extrema of $V_{BH}$ in a
general fashion. First of all, we recall that, as shown in \cite{8}, for BPS
extremal black holes it holds that
\begin{equation}
D_{i}W=\left( \partial _{i}+\partial _{i}K\right) W=0~~~~~\forall
i=1,...,n_{V},  \label{DiW}
\end{equation}
where $W$ is the holomorphic prepotential \cite{4}
\begin{equation}
W\left( z\right) =e_{\Lambda }X^{\Lambda }(z)-m^{\Lambda }F_{\Lambda }(z),~~~%
\overline{\partial }_{\overline{i}}W(z)=0,\forall \overline{i}=1,...,n_{V}.
\label{W}
\end{equation}
Here $\left( X^{\Lambda }(z),F_{\Lambda }(z)\right) $ are the holomorphic
sections of SKG. Clearly, Eq. (\ref{DiW}) can also be rewritten as
\begin{equation}
D_{i}Z=\left( \partial _{i}+\frac{1}{2}\partial _{i}K\right) Z=0~~~~~\forall
i=1,...,n_{V},  \label{DiZ}
\end{equation}
with
\begin{eqnarray}
Z\left( z,\overline{z};e,m\right)  &=&e_{\Lambda }L^{\Lambda }(z,\overline{z}%
)-m^{\Lambda }M_{\Lambda }(z,\overline{z}),~~~~\overline{D}_{\overline{i}%
}Z=0,\forall \overline{i}=1,...,n_{V},  \label{Z} \\
&&  \notag
\end{eqnarray}
where $\left( L^{\Lambda }(z,\overline{z}),M_{\Lambda }(z,\overline{z}%
)\right) =e^{K(z,\overline{z})/2}\left( X^{\Lambda }(z),F_{\Lambda
}(z)\right) $ are the K\"{a}hler-covariantly holomorphic sections of the $%
U(1)$ Hodge bundle of SKG.

Since
\begin{equation}
D_{i}V_{BH}=\partial _{i}V_{BH}=2\overline{Z}D_{i}Z+g^{j\overline{j}}\left(
D_{i}D_{j}Z\right) \overline{D}_{\overline{j}}\overline{Z},  \label{crit-pot}
\end{equation}
it is obvious that $D_{i}Z=0$ is an extremum of the potential. Moreover,
such an extremum is stable, because the Hessian of $V_{BH}$ evaluated at
this point is (strictly) positive definite \cite{1} :
\begin{eqnarray}
&&
\begin{array}{l}
\left( D_{i}D_{j}V_{BH}\right) _{BPS-extr.}=\left( \partial _{i}\partial
_{j}V_{BH}\right) _{BPS-extr.}=0, \\
\\
\\
\left( D_{i}\overline{D}_{\overline{j}}V_{BH}\right) _{BPS-extr.}=\left(
\partial _{i}\overline{\partial }_{\overline{j}}V_{BH}\right)
_{BPS-extr.}=2\left( g_{i\overline{j}}V_{BH}\right) _{BPS-extr.}= \\
\\
~~~~~~~~~~~~~~~~~~~~~~~~~~~=2\left. g_{i\overline{j}}\right|
_{BPS-extr.}\left| Z\right| _{BPS-extr.}^{2}>0,
\end{array}
\label{SUSY-crit} \\
&&  \notag
\end{eqnarray}
where the notation ``$>0$'' (``$<0$'') is clearly understood as strict
positive (negative) definiteness throughout all the present work. Notice
that the Hermiticity and (strict) positive definiteness of the Hessian of $%
V_{BH}$ at the critical supersymmetric points $D_{i}Z=0$ are due to the
Hermiticity and (strict) positive definiteness of the metric $g_{i\overline{j%
}}$ of the moduli space. If such a manifold is endowed with SKG, it holds
that
\begin{equation}
D_{i}D_{j}Z=iC_{ijk}g^{k\overline{k}}\overline{D}_{\overline{k}}\overline{Z},
\label{SKG1}
\end{equation}
trivially vanishing at $D_{i}Z=0$. If we consider the $N=2$ Lagrangian in a
black hole background \cite{5} and we denote by $\psi $ the gravitino and by
$\lambda ^{i}$ the gaugino fields, it is easy to see that such a Lagrangian
contains terms of the type
\begin{equation}
\begin{array}{l}
Z\psi \psi ; \\
\\
C_{ijk}g^{k\overline{k}}\left( \overline{D}_{\overline{k}}\overline{Z}%
\right) \lambda ^{i}\lambda ^{j}; \\
\\
\left( D_{i}Z\right) \lambda ^{i}\psi ,
\end{array}
\end{equation}
in such a way that at the BPS critical points the gaugino mass term and the $%
\lambda \psi $ term vanish, while the gravitino ``apparent mass'' term is
proportional to the value of $Z$. This is of course a consequence of the
fact that the horizon geometry at the BPS points is Bertotti-Robinson $%
AdS_{2}\times S^{2}$ (\cite{BR1}, \cite{BR2}, \cite{BR3}).

We also remark that at the BPS supersymmetric Attractor points we can also
express the black hole charges $\left( e_{\Lambda },m^{\Lambda }\right) $ in
terms of the SKG data by the so-called BPS supersymmetric Attractor
Equations \cite{8}
\begin{equation}
\left\{
\begin{array}{l}
e_{\Lambda }=2Re\left( -iZ\overline{M}_{\Lambda }\right) ; \\
\\
m^{\Lambda }=2Re\left( -iZ\overline{L}^{\Lambda }\right) ,
\end{array}
\right.   \label{SUSY-AEs}
\end{equation}
where the right-hand sides are evaluated at $D_{i}Z=0$.

\section{Non-Supersymmetric Attractors}

In \cite{1} the potential $V_{BH}$ was also considered at more general
Attractor extrema, defined by the criticity condition
\begin{equation}
\partial _{i}V_{BH}=2\overline{Z}D_{i}Z+g^{j\overline{j}}\left(
D_{i}D_{j}Z\right) \overline{D}_{\overline{j}}\overline{Z}=0~~\forall
i=1,...,n_{V}  \label{crit}
\end{equation}
and by the condition of positive definiteness of the Hessian of $V_{BH}$,
which we will, in a shorthand notation, denote as\footnote{%
By construction, $V_{BH}$ is positive definite for a (not necessarily
strictly) positive definite metric $g_{i\overline{i}}$ of the moduli space.
Consequently, the stable Attractors will necessarily be minima of such a
potential. Eqs. (\ref{crit}) and (\ref{pos}) are the general definitions of
Attractor points in the considered framework (as also recently pointed out
by \cite{GIJT1}). In the following treatment a geometry will be named
\textit{regular} if the related metric tensor is strictly positive definite.}
\begin{equation}
\left( \partial _{i}\partial _{j}V_{BH}\right) _{\partial V=0}>0.
\label{pos}
\end{equation}
It can be immediately realized that such Attractors do not necessarily
satisfy the BPS condition $D_{i}Z=0$ $\forall i=1,...,n_{V}$. In particular,
the critical points satisfying Eqs. (\ref{crit}) and (\ref{pos}) but with $%
D_{i}Z\neq 0$ for at least one index $i$ are non-supersymmetric Attractor
points, corresponding to an extremal, non-BPS black hole background with
squared mass (\cite{Kall1}, \cite{KSS1}, \cite{Tom})
\begin{eqnarray}
M_{BH}^{2} &=&\left. V_{BH}\right| _{non-BPS-extr.}=  \notag \\
&&  \notag \\
&=&\left| Z\right| _{non-BPS-extr.}^{2}+\left[ g^{i\overline{i}}\left(
D_{i}Z\right) \left( \overline{D}_{\overline{i}}\overline{Z}\right) \right]
_{non-BPS-extr.}>\left| Z\right| _{non-BPS-extr.}^{2}.  \notag \\
&&  \label{V-non-BPS}
\end{eqnarray}
Now, by substituting the SKG result (\ref{SKG1}) in the general criticity
condition (\ref{crit}), one obtains \cite{1}
\begin{equation}
\partial _{i}V_{BH}=0\Leftrightarrow 2\overline{Z}D_{i}Z+iC_{ijk}g^{j%
\overline{j}}g^{k\overline{k}}\left( \overline{D}_{\overline{j}}\overline{Z}%
\right) \left( \overline{D}_{\overline{k}}\overline{Z}\right) =0.
\label{SKG-crit}
\end{equation}
Thus, we may observe that if $D_{i}Z\neq 0$ for at least one index $i$ and $%
Z\neq 0$, then $C_{ijk}\neq 0$, i.e. the SKG rank-3 symmetric tensor will
for sure have some non-vanishing components in order for Eq. (\ref{SKG-crit}%
)\ to be satisfied.

By reconsidering the general criticality condition (\ref{crit}) (which is
independent of the eventually special Hodge-K\"{a}hler geometric structure
of the moduli space), it is easy to see that it is nothing but the general
form of the Ward identity relating the gravitino mass $Z$, the gaugino
masses $D_{i}D_{j}Z$ and the supersymmetry-breaking parameter $D_{i}Z$ in a
generic spontaneously broken supergravity theory \cite{9}. It can be recast
in the form
\begin{equation}
\left( M_{ij}h^{j}\right) _{\partial V=0}=0,  \label{EQ}
\end{equation}
with
\begin{equation}
M_{ij}\equiv D_{i}D_{j}Z+2\frac{\overline{Z}}{\left[ g^{k\overline{k}}\left(
D_{k}Z\right) \left( \overline{D}_{\overline{k}}\overline{Z}\right) \right] }%
\left( D_{i}Z\right) \left( D_{j}Z\right)
\end{equation}
and
\begin{equation}
h^{j}\equiv g^{j\overline{j}}\overline{D}_{\overline{j}}\overline{Z}.
\end{equation}
In order for Eq. (\ref{EQ}) to have a solution for non-vanishing $h^{j}$,
the $n_{V}\times n_{V}$ complex matrix $M_{ij}$ must have vanishing
determinant, so actually it has at most $n_{V}-1$ non-vanishing eigenvalues.

It is immediate to see that for $n_{V}=1$, i.e. in the one-modulus case (
\cite{GIJT1}, \cite{10}), the condition
\begin{equation}
M_{11}=0
\end{equation}
is the same as the condition $\partial _{z}V_{BH}=0$ (for $D_{z}Z\neq 0$).
By considering the SKG framework, for $n_{V}=1$ Eq. (\ref{SKG-crit}) becomes
(we introduce the scalar functions $C_{111}\equiv C\left( z,\overline{z}%
\right) \in \mathbb{C}_{0}$ and $g_{1\overline{1}}\equiv g\left( z,\overline{%
z}\right) \in \mathbb{R}_{0}^{+}$)
\begin{equation}
\partial _{z}V_{BH}=0\Leftrightarrow 2\overline{Z}D_{z}Z+iCg^{-2}\left(
\overline{D}_{\overline{z}}\overline{Z}\right) ^{2}=0,
\end{equation}
which also implies the following relation at the non-supersymmetric critical
points:
\begin{equation}
\left| D_{z}Z\right| _{non-BPS-extr.}^{2}=4\left[ g^{4}\frac{\left| Z\right|
^{2}}{\left| C\right| ^{2}}\right] _{non-BPS-extr.}.  \label{rel1}
\end{equation}
Thus, in SKG for $n_{V}=1$ Eq. (\ref{V-non-BPS}) may be written as
\begin{eqnarray}
\left. V_{BH}\right| _{non-BPS-extr.} &=&\left| Z\right|
_{non-BPS-extr.}^{2}+g^{-1}\left| D_{z}Z\right| _{non-BPS-extr.}^{2}=  \notag
\\
&&  \notag \\
&=&\left| Z\right| _{non-BPS-extr.}^{2}\left[ 1+4\left( \frac{g^{3}}{\left|
C\right| ^{2}}\right) _{non-BPS-extr.}\right] >\left| Z\right|
_{non-BPS-extr.}^{2}.  \notag \\
&&
\end{eqnarray}
Thus, we see that a general feature of the one-modulus case in SKG is that
the entropy $S_{BH,non-BPS-extr.}=\pi \left. V_{BH}\right| _{non-BPS-extr.}$
at the non-BPS, non-supersymmetric Attractors gets a multiplicative
``renormalization''
\begin{equation}
S_{BH,non-BPS-extr.}=\pi \gamma \left| Z\right| _{non-BPS-extr.}^{2},
\end{equation}
with
\begin{eqnarray}
\gamma -1 &=&4\left( \frac{g^{3}}{\left| C\right| ^{2}}\right)
_{non-BPS-extr.}>0. \\
&&  \notag
\end{eqnarray}
Consequently, the strict positivity of $\gamma -1$ and the multiplicative
``renormalization'' of the black hole entropy yield that (at least in the
considered framework) the BPS and non-BPS cases of extremal black hole
Attractors are ``discretely disjoint'' one from the other.

\section{The Hessian Matrix}

Another interesting issue we intend to address here is the study of the
condition
\begin{equation}
\frac{\partial ^{2}V_{BH}}{\partial \varphi ^{a}\partial \varphi ^{b}}%
>0~~~at~~~\frac{\partial V_{BH}}{\partial \varphi ^{a}}=0,  \label{stab}
\end{equation}
namely the condition of (strict) positive definiteness of the Hessian matrix
of $V_{BH}$ (which is nothing but the squared mass matrix of the moduli) at
the critical points of $V_{BH}$. As previously pointed out, since $V_{BH}$
is positive definite, a stable critical point (i.e. an Attractor) is
necessarily a minimum, satisfying Eq. (\ref{stab}). The $\varphi ^{a}$'s ($%
a=1,...,2n_{V}$) are real moduli, whose complexification may be given by the
unitary transformation
\begin{equation}
z^{k}\equiv \frac{\varphi ^{2k-1}+i\varphi ^{2k}}{\sqrt{2}},~~k=1,...,n_{V}.
\label{unit-transf}
\end{equation}
From the (strict) positive definiteness of the metric tensor $g_{i\overline{j%
}}$ in a regular moduli space, we have seen above that the condition (\ref
{stab}) is automatically satisfied at the BPS supersymmetric Attractor
points (defined by $D_{i}Z=0$ $\forall i$). Of course, for $D_{i}Z\neq 0$
for at least some index $i$, this is no longer true, and a more detailed
analysis of (\ref{stab}) is needed.

In the complex basis, the $2n_{V}\times 2n_{V}$ Hessian of $V_{BH}$ reads
\begin{equation}
H_{\widehat{i}\widehat{j}}^{V_{BH}}\equiv \left(
\begin{array}{ccc}
D_{i}D_{j}V_{BH} &  & D_{i}\overline{D}_{\overline{j}}V_{BH} \\
&  &  \\
D_{j}\overline{D}_{\overline{i}}V_{BH} &  & \overline{D}_{\overline{i}}%
\overline{D}_{\overline{j}}V_{BH}
\end{array}
\right) =\left(
\begin{array}{ccc}
\mathcal{M}_{ij} &  & \mathcal{N}_{i\overline{j}} \\
&  &  \\
\overline{\mathcal{N}}_{j\overline{i}} &  & \overline{\mathcal{M}}_{%
\overline{i}\overline{j}}
\end{array}
\right) ,
\end{equation}
where the hatted indices $\hat{\imath}$ and $\hat{\jmath}$ may be
holomorphic or anti-holomorphic.

In the SKG framework, one gets
\begin{eqnarray}
&&
\begin{array}{l}
\mathcal{M}_{ij}\equiv D_{i}D_{j}V_{BH}=D_{j}D_{i}V_{BH}= \\
\\
=4i\overline{Z}C_{ijk}g^{k\overline{k}}\left( \overline{D}_{\overline{k}}%
\overline{Z}\right) +i\left( D_{j}C_{ikl}\right) g^{k\overline{k}}g^{l%
\overline{l}}\left( \overline{D}_{\overline{k}}\overline{Z}\right) \left(
\overline{D}_{\overline{l}}\overline{Z}\right) ;
\end{array}
\\
&&  \notag \\
&&  \notag \\
&&
\begin{array}{l}
\mathcal{N}_{i\overline{j}}\equiv D_{i}\overline{D}_{\overline{j}}V_{BH}=%
\overline{D}_{\overline{j}}D_{i}V_{BH}= \\
\\
=2\left[ g_{i\overline{j}}\left| Z\right| ^{2}+\left( D_{i}Z\right) \left(
\overline{D}_{\overline{j}}\overline{Z}\right) +g^{l\overline{n}}C_{ikl}%
\overline{C}_{\overline{j}\overline{m}\overline{n}}g^{k\overline{k}}g^{m%
\overline{m}}\left( \overline{D}_{\overline{k}}\overline{Z}\right) \left(
D_{m}Z\right) \right] .
\end{array}
\\
&&  \notag
\end{eqnarray}
Thus, $\mathcal{M}^{T}=\mathcal{M}$ and $\mathcal{N}^{\dag }=\mathcal{N}$.
Here we just note that for symmetric special K\"{a}hler-Hodge moduli spaces $%
D_{j}C_{ikl}=0$, and that the tensor $g^{l\overline{n}}C_{ikl}\overline{C}_{%
\overline{j}\overline{m}\overline{n}}$ can be rewritten in terms of the
Riemann-Christoffel tensor by using the so-called ``SKG constraints'' (see
e.g. \cite{4}) :
\begin{equation}
g^{l\overline{n}}C_{ikl}\overline{C}_{\overline{j}\overline{m}\overline{n}%
}=g_{i\overline{j}}g_{k\overline{m}}+g_{i\overline{m}}g_{k\overline{j}}-R_{i%
\overline{j}k\overline{m}}.  \label{SKG-constr}
\end{equation}
Note that the given expression for $\mathcal{M}_{ij}$ is actually symmetric,
since through the constraints (\ref{SKG-constr}) the Bianchi identities for $%
R_{i\overline{j}k\overline{m}}$ imply $D_{[j}C_{i]kl}=0$, with square
brackets denoting antisymmetrization of enclosed indices.

At the critical points of $V_{BH}$, the K\"{a}hler-covariant derivatives may
be substituted by the flat derivatives, and the Hessian becomes
\begin{equation}
\left. H_{\widehat{i}\widehat{j}}^{V_{BH}}\right| _{\partial V=0}\equiv
\left(
\begin{array}{ccc}
\partial _{i}\partial _{j}V_{BH} &  & \partial _{i}\overline{\partial }_{%
\overline{j}}V_{BH} \\
&  &  \\
\partial _{j}\overline{\partial }_{\overline{i}}V_{BH} &  & \overline{%
\partial }_{\overline{i}}\overline{\partial }_{\overline{j}}V_{BH}
\end{array}
\right) _{\partial V=0}=\left(
\begin{array}{ccc}
\mathcal{M} &  & \mathcal{N} \\
&  &  \\
\mathcal{N} &  & \overline{\mathcal{M}}
\end{array}
\right) _{\partial V=0}.
\end{equation}
Instead, in the real basis the Hessian is a $2n_{V}\times 2n_{V}$ real
symmetric matrix, which may be decomposed as follows:
\begin{equation}
\frac{\partial ^{2}V_{BH}}{\partial \varphi ^{a}\partial \varphi ^{b}}%
=\left(
\begin{array}{ccc}
\mathcal{A} &  & \mathcal{C} \\
&  &  \\
\mathcal{C}^{T} &  & \mathcal{B}
\end{array}
\right) ,
\end{equation}
with $\mathcal{A}$ and $\mathcal{B}$ being $n_{V}\times n_{V}$ real
symmetric matrices ($\mathcal{A}^{T}=\mathcal{A}$, $\mathcal{B}^{T}=\mathcal{%
B}$). By relating the real and complex basis of the special K\"{a}hler
moduli space $\mathcal{V}$ through the unitary transformation (\ref
{unit-transf}), the relation between the complex matrices $\mathcal{M}$, $%
\mathcal{N}$ and $\mathcal{A}$, $\mathcal{B}$ is given by
\begin{equation}
\left\{
\begin{array}{l}
\mathcal{M}=\left( \mathcal{A}-\mathcal{B}\right) +i\left( \mathcal{C}+%
\mathcal{C}^{T}\right) ; \\
\\
\mathcal{N}=\left( \mathcal{A}+\mathcal{B}\right) +i\left( \mathcal{C}^{T}-%
\mathcal{C}\right) ,
\end{array}
\right.
\end{equation}
or its inverse
\begin{equation}
\left\{
\begin{array}{l}
\mathcal{A}=\frac{1}{2}\left( Re\mathcal{M}+Re\mathcal{N}\right) ; \\
\\
\mathcal{B}=\frac{1}{2}\left( Re\mathcal{N}-Re\mathcal{M}\right) ; \\
\\
\mathcal{C}=\frac{1}{2}\left( Im\mathcal{M}-Im\mathcal{N}\right) .
\end{array}
\right.
\end{equation}
It is worth noticing that the unitary transformation (\ref{unit-transf}) may
be represented by a unitary ($U(2n_{V})$) matrix $\mathcal{U}$, acting as
follows ($\mathcal{U}^{-1}=\mathcal{U}^{\dag }$):
\begin{equation}
H_{\mathbb{R}}^{V_{BH}}=\mathcal{U}^{-1}H_{\mathbb{C}}^{V_{BH}}\left(
\mathcal{U}^{T}\right) ^{-1},
\end{equation}
or equivalently
\begin{equation}
H_{\mathbb{C}}^{V_{BH}}=\mathcal{U}H_{\mathbb{R}}^{V_{BH}}\mathcal{U}^{T},
\end{equation}
where the subscripts ``$\mathbb{R}$'' and ``$\mathbb{C}$'' denote the real
and complex parametrization of the Hessian $H^{V_{BH}}$, respectively.

Now, it should be recalled that the properties of SKG yield that at a
generic point in the Hodge-K\"{a}hler moduli space $\mathcal{V}$, the
following relation holds true \cite{Attr-Black} :
\begin{equation}
\mathcal{P}-i\Omega \mathbf{M}\left( \mathbf{N}\right) \mathcal{P}=-2iZ%
\overline{V}-2iG^{j\overline{j}}\left( \overline{D}_{\overline{j}}\overline{Z%
}\right) \left( D_{j}V\right) ,  \label{fund-rela}
\end{equation}
where $\mathcal{P}$ and $V$ respectively are the $Sp\left( 2n_{V}\right) $%
-covariant vectors of black hole charges and of the K\"{a}hler-covariantly
holomorphic sections of the SKG:
\begin{equation}
\mathcal{P}\equiv \left(
\begin{array}{c}
m^{\Lambda } \\
\\
e_{\Lambda }
\end{array}
\right) ,~~~~~V\left( z,\overline{z}\right) \equiv \left(
\begin{array}{c}
L^{\Lambda }\left( z,\overline{z}\right)  \\
\\
M_{\Lambda }\left( z,\overline{z}\right)
\end{array}
\right) ,~~\overline{D}_{\overline{i}}V\left( z,\overline{z}\right) =0,
\end{equation}
and $\Omega $ is the $\left( 2n_{V}+2\right) $-dim. symplectic metric:
\begin{equation}
\Omega \equiv \left(
\begin{array}{cc}
0 & -\mathbb{I} \\
\mathbb{I} & 0
\end{array}
\right)
\end{equation}
($\mathbb{I}$ stands for the $\left( n_{V}+1\right) $-dim. unit matrix). $%
\mathbf{M}\left( \mathbf{N}\right) $ is a real $\left( 2n_{V}+2\right) $%
-dim. square matrix, defined as (\cite{4}, \cite{8})
\begin{eqnarray}
&&
\begin{array}{l}
\mathbf{M}\left( Re\left( \mathbf{N}\right) ,Im\left( \mathbf{N}\right)
\right) \equiv  \\
\\
\\
\equiv \left(
\begin{array}{cccc}
Im\left( \mathbf{N}\right) +Re\left( \mathbf{N}\right) \left( Im\left(
\mathbf{N}\right) \right) ^{-1}Re\left( \mathbf{N}\right)  &  & ~~~~~~~~ &
-Re\left( \mathbf{N}\right) \left( Im\left( \mathbf{N}\right) \right) ^{-1}
\\
&  &  &  \\
&  &  &  \\
-\left( Im\left( \mathbf{N}\right) \right) ^{-1}Re\left( \mathbf{N}\right)
&  &  & \left( Im\left( \mathbf{N}\right) \right) ^{-1}
\end{array}
\right) ,
\end{array}
\notag \\
&&  \notag \\
&&
\end{eqnarray}
where $\mathbf{N}_{\Lambda \Sigma }$ is a $\left( n_{V}+1\right) \times
\left( n_{V}+1\right) $ moduli-dependent complex symmetric matrix defined by
the fundamental Ans\"{a}tze of SKG (see e.g. \cite{4}):
\begin{eqnarray}
M_{\Lambda }\left( z,\overline{z}\right)  &\equiv &\mathbf{N}_{\Lambda
\Sigma }\left( z,\overline{z}\right) L^{\Sigma }\left( z,\overline{z}\right)
; \\
&&  \notag \\
D_{i}M_{\Lambda }\left( z,\overline{z}\right)  &\equiv &\overline{\mathbf{N}}%
_{\Lambda \Sigma }\left( z,\overline{z}\right) D_{i}L^{\Sigma }\left( z,%
\overline{z}\right) .
\end{eqnarray}
By taking the real part\footnote{%
Actually, the imaginary part of the SKG relation (\ref{fund-rela}) is a new
identity satisfied by the central charge $Z$.} of the SKG identity (\ref
{fund-rela}) and evaluating it at the critical points of $V_{BH}$
(satisfying Eq. (\ref{SKG-crit})), one gets the general form of the extremal
black hole Attractor Eqs. (\cite{FBC}, \cite{Kall1}, \cite{KSS1}; see also
\cite{Tom}) :
\begin{equation}
\mathcal{P}=2\left\{ Re\left[ -iZ\overline{V}+\frac{1}{2Z}\overline{C}_{%
\overline{i}\overline{j}\overline{k}}g^{i\overline{i}}g^{j\overline{j}}g^{k%
\overline{k}}\left( D_{j}Z\right) \left( D_{k}Z\right) \left( D_{i}V\right)
\right] \right\} _{\partial V=0}.  \label{gen-AEs}
\end{equation}
Notice that at the BPS supersymmetric extremal black hole Attractors
(satisfying Eq. (\ref{DiW}) or equivalently Eq. (\ref{DiZ})) Eqs. (\ref
{gen-AEs}) consistently reduce to the well-known supersymmetric extremal
black hole Attractor Eqs. (\ref{SUSY-AEs}).

Let us now reconsider the one-modulus case. For $n_{V}=1$, the
moduli-dependent matrices $\mathcal{A}$, $\mathcal{B}$, $\mathcal{C}$, $%
\mathcal{M}$ and $\mathcal{N}$ introduced above are simply scalar functions.
In particular, $\mathcal{N}$ is real, since $\mathcal{C}$ trivially
satisfies $\mathcal{C}=\mathcal{C}^{T}$. The stability condition (\ref{stab}%
) in this case reads
\begin{equation}
\frac{\partial ^{2}V_{BH}}{\partial \varphi ^{a}\partial \varphi ^{b}}%
>0~~~at~~~\frac{\partial V_{BH}}{\partial \varphi ^{a}}=0,
\end{equation}
with $a,b=1,2$, and it may be rewritten as
\begin{equation}
\left. \mathcal{N}\right| _{\partial V=0}>\left| \mathcal{M}\right|
_{\partial V=0},  \label{stab-1}
\end{equation}
where
\begin{eqnarray}
\mathcal{N} &\equiv &D_{z}\overline{D}_{\overline{z}}V_{BH}=\overline{D}_{%
\overline{z}}D_{z}V_{BH}=2\left[ g\left| Z\right| ^{2}+\left| D_{z}Z\right|
^{2}+\left| C\right| ^{2}g^{-3}\left| D_{z}Z\right| ^{2}\right] ; \\
&&  \notag \\
\mathcal{M} &\equiv &D_{z}D_{z}V_{BH}=4i\overline{Z}Cg^{-1}\left( \overline{D%
}_{\overline{z}}\overline{Z}\right) +i\left( D_{z}C\right) g^{-2}\left(
\overline{D}_{\overline{z}}\overline{Z}\right) ^{2}.
\end{eqnarray}
Once again, the 1-modulus stability condition (\ref{stab-1}) is
automatically satisfied for $D_{z}Z=0$, i.e. at BPS supersymmetric
Attractors. Instead, by evaluating the functions $\mathcal{N}$ and $\mathcal{%
M}$ at the non-BPS, non-supersymmetric critical points of $V_{BH}$ and using
the relation (\ref{rel1}), one finally gets
\begin{equation}
\left. \mathcal{N}\right| _{non-BPS-extr.}=2\left[ \left| D_{z}Z\right|
^{2}\left( 1+\frac{5}{4}\left| C\right| ^{2}g^{-3}\right) \right]
_{non-BPS-extr.};
\end{equation}
\begin{eqnarray}
&&
\begin{array}{l}
\left| \mathcal{M}\right| _{non-BPS-extr.}^{2}= \\
\\
=4\left\{ \left| D_{z}Z\right| ^{4}\left[ \left| C\right| ^{4}g^{-6}+\frac{1%
}{4}g^{-4}\left| D_{z}C\right| ^{2}+2g^{-3}Re\left[ C\left( \overline{D}_{%
\overline{z}}\overline{C}\right) \left( \overline{D}_{\overline{z}}ln%
\overline{Z}\right) \right] \right] \right\} _{non-BPS-extr.}.
\end{array}
\notag \\
&&
\end{eqnarray}
Therefore, Eq. (\ref{stab-1}) yields that the stability condition for
non-BPS, non-supersymmetric extremal black hole Attractors in the $n_{V}=1$
SKG framework reads
\begin{gather}
\left. \mathcal{N}\right| _{non-BPS-extr.}>\left| \mathcal{M}\right|
_{non-BPS-extr.}; \\
\Updownarrow   \notag \\
1+\frac{5}{4}\left( \left| C\right| ^{2}g^{-3}\right) _{non-BPS-extr.}>
\notag \\
\notag \\
>\sqrt{\left[ \left| C\right| ^{4}g^{-6}+\frac{1}{4}g^{-4}\left|
D_{z}C\right| ^{2}+2g^{-3}Re\left[ C\left( \overline{D}_{\overline{z}}%
\overline{C}\right) \left( \overline{D}_{\overline{z}}ln\overline{Z}\right)
\right] \right] _{non-BPS-extr.}}.  \label{stab-1-1} \\
\notag
\end{gather}
One may immediately notice that Eq. (\ref{stab-1-1}) is surely satisfied for
$D_{z}C=0$, corresponding to 1-dimensional complex symmetric special K\"{a}%
hler spaces \cite{11}, but of course this is not the only possibility. In
the general case, a new fundamental data, $D_{z}C$ turns out to play a
crucial role in determining the stability of non-supersymmetric Attractors.

\section{K\"{a}hler Gauges and Entropic Functional}

\bigskip In this last Section we discuss some properties of the ``black hole
effective potential'' function $V_{BH}$ given by Eq. (\ref{VBH}).

First of all, we observe that the ``central charge'' function $Z$ given by
Eq. (\ref{Z}) is a section of the $U(1)$ Hodge bundle over the special K\"{a}%
hler manifold $\mathcal{V}$; indeed, under a general K\"{a}hler gauge
transformation
\begin{equation}
K\left( z,\overline{z}\right) \longrightarrow K\left( z,\overline{z}\right)
+f\left( z\right) +\overline{f}\left( \overline{z}\right)
\end{equation}
the holomorphic sections (having K\"{a}hler weights $\left( 2,0\right) $)
transform as
\begin{equation}
X^{\Lambda }\left( z\right) \longrightarrow X^{\Lambda }\left( z\right)
e^{-f(z)},~~~~~F_{\Lambda }\left( z\right) \longrightarrow F_{\Lambda
}\left( z\right) e^{-f(z)},
\end{equation}
and therefore, Eq. (\ref{Z}) yields that $Z$ transforms by a pure phase:
\begin{equation}
Z\left( z,\overline{z};e,m\right) \longrightarrow Z\left( z,\overline{z}%
;e,m\right) e^{-\frac{1}{2}\left( f(z)-\overline{f}(\overline{z})\right)
}=Z\left( z,\overline{z};e,m\right) e^{-iIm\left[ f(z)\right] }.
\end{equation}
By recalling Eq. (\ref{W}), it is instead clear that the ``holomorphic
central charge'' function $W\left( z;e,m\right) $, like the holomorphic
sections $X^{\Lambda }\left( z\right) $ and $F_{\Lambda }\left( z\right) $,
is a section of the line (Hodge) bundle \cite{5}, since it transforms as
\begin{equation}
W\left( z;e,m\right) \longrightarrow W\left( z;e,m\right) e^{-f(z)}.
\end{equation}

If we introduce the K\"{a}hler gauge-invariant real function \cite{12} (well
defined if\linebreak\ $\left| W\right| \left( z,\overline{z};e,m\right) \neq
0$)
\begin{gather}
G\left( z,\overline{z};e,m\right) \equiv K\left( z,\overline{z}\right) +ln%
\left[ \left| W\right| ^{2}\left( z,\overline{z};e,m\right) \right]  \\
\Downarrow   \notag \\
e^{G\left( z,\overline{z};e,m\right) }=e^{K\left( z,\overline{z}\right)
}\left| W\right| ^{2}\left( z,\overline{z};e,m\right) =\left| Z\right|
^{2}\left( z,\overline{z};e,m\right) ,
\end{gather}
it is easy to check that the potential $V_{BH}$ given by Eq. (\ref{VBH}) may
be rewritten as
\begin{equation}
V_{BH}\left( z,\overline{z};e,m\right) =e^{G\left( z,\overline{z};e,m\right)
}\left\{ 1+g^{i\overline{i}}\left( z,\overline{z}\right) \left[ \partial
_{i}G\left( z,\overline{z};e,m\right) \right] \left[ \overline{\partial }_{%
\overline{i}}G\left( z,\overline{z};e,m\right) \right] \right\} .
\label{VBH2}
\end{equation}
Thus, for a given black hole charge configuration $\left( e_{\Lambda
},m^{\Lambda }\right) $ and non-vanishing $Z$, an equivalent expression of
the definitions (\ref{DiW}) and (\ref{DiZ}) of the BPS, supersymmetric
Attractor points reads as follows:
\begin{equation}
\partial _{i}G\left( z,\overline{z};e,m\right) =0,~~~\forall i=1,...,n_{V};
\label{diG=0}
\end{equation}
at such points $V_{BH}$ reduces to
\begin{equation}
\left. V_{BH}\right| _{\partial G=0}=\left. e^{G}\right| _{\partial
G=0}=\left| Z\right| _{\partial G=0}^{2}.  \label{VV}
\end{equation}
Moreover, at supersymmetric Attractors Eq. (\ref{SUSY-crit}) yields
\begin{equation}
\begin{array}{l}
\left( D_{i}D_{j}V_{BH}\right) _{\partial G=0}=\left( \partial _{i}\partial
_{j}V_{BH}\right) _{\partial G=0}=0, \\
\\
\\
\left( D_{i}\overline{D}_{\overline{j}}V_{BH}\right) _{\partial G=0}=\left(
\partial _{i}\overline{\partial }_{\overline{j}}V_{BH}\right) _{\partial
G=0}=2\left( g_{i\overline{j}}V_{BH}\right) _{\partial G=0}= \\
\\
~~~~~~~~~~~~~~~~~~~~~\ =2\left( g_{i\overline{j}}e^{G}\right) _{\partial
G=0}.
\end{array}
\label{SUSY-crit2}
\end{equation}
Thus, one obtains once again that for a positive definite metric tensor $%
\left. g_{i\overline{j}}\right| _{\partial G=0}$ (as it should be in regular
SKG) all the critical points of $V_{BH}$ satisfying the condition (\ref
{diG=0}) are actually minima of $V_{BH}$, corresponding to BPS Attractors.

Let us now consider the relevance of the above results in relation to the
entropic functional given by Gukov, Saraikin and Vafa (GSV) in Eq. (3.1) of
\cite{GSV}. It reads
\begin{equation}
\mathcal{S}_{GSV}=\frac{\pi }{4}e^{-K}.
\end{equation}
Since it holds that
\begin{equation}
e^{-K}=\left. e^{-G}\right| _{W=1}=\left. \left| Z\right| ^{-2}\right|
_{W=1},
\end{equation}
it is clear that $\mathcal{S}_{GSV}$ is $\frac{\pi }{4}$ times $e^{-G}$ in a
K\"{a}hler gauge in which $W=1$:
\begin{equation}
\mathcal{S}_{GSV}=\frac{\pi }{4}e^{-K}=\frac{\pi }{4}\left. e^{-G}\right|
_{W=1}=\frac{\pi }{4}\left. \left| Z\right| ^{-2}\right| _{W=1}.
\end{equation}
Therefore, due to the K\"{a}hler gauge-invariance of $G$, in a regular SKG
framework with non-vanishing $Z$, the second K\"{a}hler-covariant
derivatives of $\mathcal{S}_{GSV}$ at the BPS Attractors read as follows:
\begin{eqnarray}
\left( D_{i}D_{j}\mathcal{S}_{GSV}\right) _{\partial G=0} &=&\left(
D_{i}D_{j}e^{-G}\right) _{\partial G=0}=\left( \partial _{i}\partial
_{j}e^{-G}\right) _{\partial G=0}=  \notag \\
&&  \notag \\
&=&-\left( \partial _{i}\partial _{j}G\right) _{\partial G=0}\left.
e^{-G}\right| _{\partial G=0}=0;  \label{final-1} \\
&&  \notag
\end{eqnarray}
\begin{eqnarray}
\left( \overline{D}_{\overline{i}}D_{j}\mathcal{S}_{GSV}\right) _{\partial
G=0} &=&\left( \overline{D}_{\overline{i}}D_{j}e^{-G}\right) _{\partial
G=0}=\left( \overline{\partial }_{\overline{i}}\partial _{j}e^{-G}\right)
_{\partial G=0}=  \notag \\
&&  \notag \\
&=&-\left( \partial _{i}\overline{\partial }_{\overline{j}}G\right)
_{\partial G=0}\left. e^{-G}\right| _{\partial G=0}<0,  \label{final-2} \\
&&  \notag
\end{eqnarray}
where the results of \cite{1}, concerning the strict positive definiteness
of the Hessians of $V_{BH}$ and $\left| Z\right| $ at the supersymmetric
Attractors, have been used; they imply that, in the assumptions made above,
at $\partial _{i}G=0$ $\forall i=1,...,n_{V}$ the Hessian of the K\"{a}hler
gauge-invariant function $G$ is strictly positive definite.

Summarizing, for a given black hole charge configuration $\left( e_{\Lambda
},m^{\Lambda }\right) $, in the regular SKG and for non-vanishing $Z$, Eqs. (%
\ref{final-1}) and (\ref{final-2}) yield that all the moduli configurations
satisfying the BPS condition (\ref{diG=0}) are maxima of the GSV entropic
functional $\mathcal{S}_{GSV}$ or of its rigorous K\"{a}hler gauge-invariant
completion $e^{-G}$. Such a result confirms the recent observation of \cite
{Fiol}.

Finally, it is worth pointing out that it is hard to give a meaning to the
Hessian of $\mathcal{S}_{GSV}$ or of its rigorous K\"{a}hler gauge-invariant
completion $e^{-G}$ at non-BPS, non-supersymmetric Attractor points. Indeed,
such points are extrema of the whole ``black hole potential'' function $%
V_{BH}$ given by Eqs. (\ref{VBH}) and (\ref{VBH2}), but not of (the inverse
of) its BPS, supersymmetric part, proportional to $\mathcal{S}_{GSV}$ and $%
e^{-G}$. \vspace{20pt}

\textbf{Acknowledgments}

The work of S.B. has been supported in part by the European Community Human
Potential Program under contract MRTN-CT-2004-005104 ``Constituents,
fundamental forces and symmetries of the universe''.\textbf{\ }The work of
S.F.~has been supported in part by the European Community Human Potential
Program under contract MRTN-CT-2004-005104 ``Constituents, fundamental
forces and symmetries of the universe'', in association with INFN Frascati
National Laboratories and by D.O.E.~grant DE-FG03-91ER40662, Task C. The
work of A.M. has been supported by a Junior Grant of the ``Enrico Fermi''
Center, Rome, in association with INFN Frascati National Laboratories.

\end{document}